\documentclass[aps,pra,10pt,twocolumn,floatfix,superscriptaddress]{revtex4-2}

\usepackage[T1]{fontenc}
\usepackage[colorlinks=true,urlcolor=blue,linkcolor=blue,citecolor=blue]{hyperref}
\usepackage{graphicx,overpic,orcidlink,amssymb,bm,physics}

\newcommand{\orcMarco}{\orcidlink{0000-0002-3215-3453}}
\newcommand{\orcJuanjo}{\orcidlink{0000-0001-8993-4624}}
\newcommand{\orcDave}{\orcidlink{0000-0001-8546-9075}}
\newcommand{\orcLubi}{\orcidlink{0000-0002-2636-9936}}
\newcommand{\affA}{\affiliation{Quantinuum, Partnership House, Carlisle Place, London SW1P 1BX, United Kingdom}}
\newcommand{\affB}{\affiliation{Instituto de F\'{i}sica Fundamental, IFF-CSIC, Calle Serrano 113b, Madrid 28006, Spain}}
\newcommand{\affC}{\affiliation{Quantinuum, 303 S.\ Technology Ct., Broomfield, Colorado 80021, USA}}

\begin{document}

\title{Efficient quantum state preparation of multivariate functions using tensor networks}

\author{Marco Ballarin\orcMarco}\email{marco.ballarin@quantinuum.com}\affA
\author{Juan Jos\'{e} Garc\'{i}a-Ripoll\orcJuanjo}\affB
\author{David Hayes\orcDave}\affC
\author{Michael Lubasch\orcLubi}\email{michael.lubasch@quantinuum.com}\affA

\date{November 19, 2025}

\begin{abstract}
For the preparation of high-dimensional functions on quantum computers, we introduce tensor network algorithms that are efficient with regard to dimensionality, optimize circuits composed of hardware-native gates and take gate errors into account during the optimization.
To avoid the notorious barren plateau problem of vanishing gradients in the circuit optimization, we smoothly transform the circuit from an easy-to-prepare initial function into the desired target function.
We show that paradigmatic multivariate functions can be accurately prepared such as, by numerical simulations, a 17-dimensional Gaussian encoded in the state of 102 qubits and, through experiments, a 9-dimensional Gaussian realized using 54 qubits on Quantinuum's H2 quantum processor.
\end{abstract}

\maketitle

\textit{Introduction}---Quantum state preparation (QSP) is a necessary component of several quantum algorithms.
For example, in Hamiltonian simulation, QSP is required to prepare the initial state that is propagated according to the time-dependent Schr\"{o}dinger equation~\cite{nielsen2010quantum,dalzell2025quantum}.
And quantum linear systems algorithms need QSP to create the vector of constant terms of the linear system on a quantum computer~\cite{aaronson2015read,morales2025quantumlinearsolverssurvey}.
Numerous additional examples can be found, with applications in quantum chemistry~\cite{cao2019quantum,mcardle2020quantum}, computational fluid dynamics~\cite{jaksch2023,tennie2025quantum}, structural mechanics~\cite{liu2024towards} and finance~\cite{orus2019,herman2023quantum}, among others.
Optimized implementations of these algorithms necessitate that QSP is as efficient as possible.

While the computational cost of QSP scales exponentially with the number of qubits in general~\cite{vartiainen_2004,mottonen_2005,zhang_2022,sun_2023}, efficient approaches exist for the preparation of specific functions~\cite{grover2002creating,kitaev2008wavefunction,lubasch_2020,holmes_efficient_2020,rattew2021efficient,plekhanov_2022,rattew2022preparing,melnikov_quantum_2023,akhalwaya_2023,marin2023approx,jumade_data_2023,gonzalezConde2024efficientquantum,ben-dov_2024,wright2024noisy,bohun_scalable_2024,green_quantum_2025,rosenkranz_2025,sano_quantum_2025}.
Among these, tensor network (TN) algorithms~\cite{silvi2019,orus2019tensor} form a particularly versatile toolbox to efficiently encode a large class of functions~\cite{khoromskij2011d,oseledets_2013,garcia-ripoll_2021} and optimize shallow parameterized quantum circuits (PQCs)~\cite{lubasch_2020,holmes_efficient_2020,plekhanov_2022,melnikov_quantum_2023,akhalwaya_2023,jumade_data_2023,ben-dov_2024,bohun_scalable_2024,green_quantum_2025,sano_quantum_2025}.
In the context of function approximation, a useful TN method is the so-called tensor cross interpolation (TCI) which can efficiently encode a black-box function in a tree TN (TTN)~\cite{oseledets_tt-cross_2010,nunez_fernandez_learning_2022,ritter_quantics_2024,tindall_compressing_2024,fernandez_learning_2025}.
The TTN state, in turn, has emerged as a promising variational ansatz to represent high-dimensional functions as its hierarchical structure can naturally capture non-trivial correlations between many variables~\cite{tindall_compressing_2024}.
Recent studies have demonstrated the potential of TN techniques for multivariate QSP, particularly for low-dimensional functions~\cite{pereira2024encoding} and structured multivariate distributions~\cite{manabe_state_2024}.
However, previous approaches use a mathematical mapping of TN representations to quantum circuits~\cite{schoen2005,schoen2007} that is efficient in theory but can lead to deep quantum circuits that are not optimized for current hardware.

\begin{figure}
\centering
\includegraphics[width=\columnwidth]{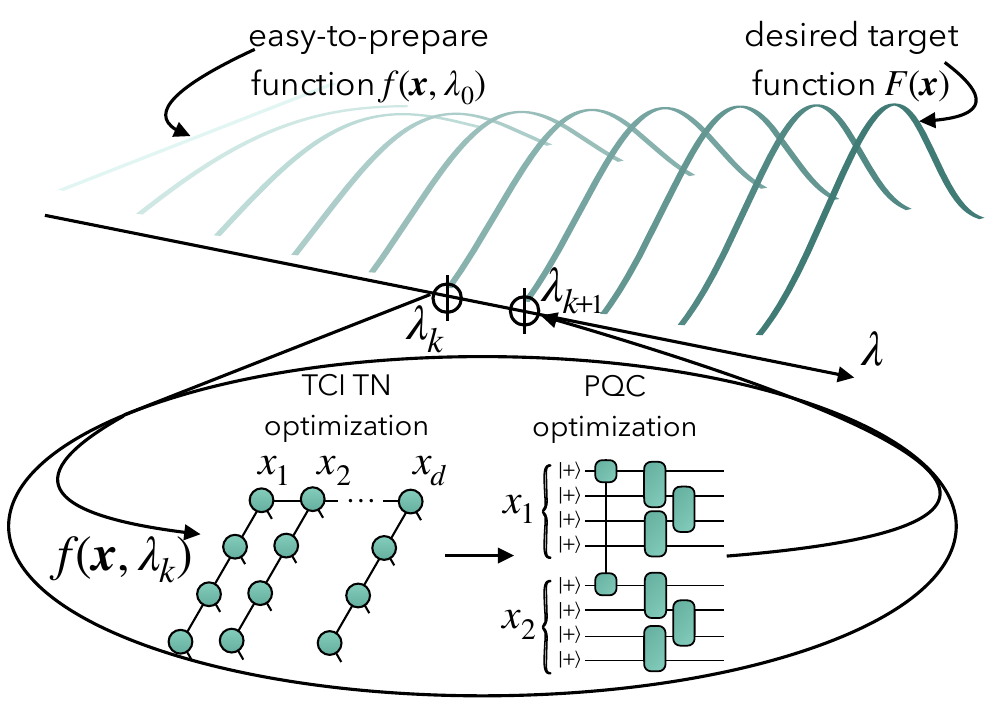}
\caption{\label{fig:1}
The proposed IQSP algorithm.
To prepare a desired target function $F(\bm{x})$ on a quantum computer, IQSP uses a sequence of smoothly connected functions $f(\bm{x}, \lambda_{k})$ where $k \in \{0, 1, \dots, K\}$ and $f(\bm{x}, \lambda_{0})$ is easy to prepare and $f(\bm{x}, \lambda_{K}) = F(\bm{x})$.
For each value of $\lambda_{k}$, IQSP first creates a TN representation of $f(\bm{x}, \lambda_{k})$ using TCI and then optimizes a PQC to maximize the fidelity with that TN.
The IQSP algorithm warm-starts the training for each $k > 0$ using the optimal parameters from the preceding step $k-1$, which can ensure large gradients throughout the circuit optimization.
Shown are a comb TN for $d$ variables as well as a PQC for $d = 2$, as they are used in the paper.
Here $\ket{+} = (\ket{0}+\ket{1}) / \sqrt{2}$ and the green circles/boxes represent variational tensors/unitaries.
}
\end{figure}

In this paper, we present efficient classical algorithms for the variational optimization of PQCs to realize multivariate QSP in a hardware-efficient way.
We achieve this by optimizing hardware-native quantum gates and incorporating realistic gate errors in the optimization.
A significant obstacle to the efficiency of PQC optimization is the barren plateau problem of vanishing gradients~\cite{mcclean_2018, cerezo_2021, uvarov_2021, zhao_2021, patti_2021, ortizmarrero_2021, wang_2021, holmes_2022, larocca_2022, cerveromartin_2023}.
To circumvent this problem, we introduce interpolative QSP (IQSP) which prepares a certain desired function by iteratively performing QSP for a sequence of functions that smoothly interpolate between an easy-to-prepare initial function and the desired final function.
The IQSP protocol warm-starts the optimization of each iteration using the optimized PQC of the previous iteration and, by doing so, can guarantee that large gradients are encountered during the optimization.
The procedure is illustrated in Fig.~\ref{fig:1} and explained in the corresponding figure caption.
Throughout the paper, for the function approximations, we use comb TNs (CTN) and comb-like PQC architectures of the structures shown in Fig.~\ref{fig:1}, because CTNs can efficiently represent interesting multivariate functions~\cite{tindall_compressing_2024}.
We emphasize that IQSP can employ other TNs and PQCs.

In the context of preparing high-dimensional Gaussians, we numerically demonstrate that IQSP can indeed avoid the barren plateau problem and successfully train circuits for up to 17 dimensions and 102 qubits.
Furthermore, we show that knowledge about gate errors can be utilized during the IQSP optimization such that the resulting circuits can be optimal also with respect to certain types of experimental noise.
With the help of the noise-aware optimization, we prepare an accurate representation of a 9-dimensional Gaussian on Quantinuum's H2-2 trapped-ion device~\cite{moses2023race,decross2025} using $255$ two-qubit and $564$ single-qubit gates.
In the End Matter, we also benchmark the performance of IQSP on the preparation of a $2$-dimensional Ricker wavelet~\cite{ricker1953form} and a Student's t-distribution~\cite{student_1908}:
Compared with the important proposal~\cite{rosenkranz_2025}, we show that IQSP can reduce the two-qubit gate count by more than one order of magnitude.

\textit{Background}---Our goal is to encode a function $f(\bm{x})$ defined on a $d$-dimensional domain $\bigotimes_{i=1}^d [0, 1) \subset \mathbb{R}^d$, i.e.\ with $\bm{x} = (x_1, \dots, x_d)$ and $x_i \in [0,1)$, in a quantum state $\ket{f}$.
Each variable $x_i$ takes on values from a grid of $2^{n_{\text{x}}}$ equidistant grid points that we identify with bitstrings via the binary representation:
\begin{align}\label{eq:discretization}
 x_i = \sum_{\alpha=1}^{n_{\text{x}}} x_{i,\alpha} 2^{-\alpha}, \quad x_{i,\alpha}\in\{0,1\},
\end{align}
such that $|x_{i}\rangle = |x_{i,1},\; \dots,\; x_{i,n_{\text{x}}}\rangle$.
Then the function is represented by the normalized quantum state of $n = d n_{\text{x}}$ qubits:
\begin{align}\label{eq:state}
 f(\bm{x}) &\rightarrow \ket{f} = \frac{1}{\mathcal{N}} \sum_{\bm{x}} f(\bm{x}) \ket{\bm{x}},
\end{align}
where $\mathcal{N} = \sqrt{\sum_{\bm{x}} |f(\bm{x})|^2}$ is the normalization factor and $\ket{\bm{x}} = \bigotimes_{i=1}^d \ket{x_i}$.

The TCI algorithm can efficiently compute a TN approximation of such a quantum state.
It was first developed for the famous matrix product state (MPS) ansatz~\cite{oseledets_tt-cross_2010} and later generalized to TTNs~\cite{tindall_compressing_2024}.
Given a $d$-variate function $f(\bm{x})$, the amplitude-encoded quantum state can be compressed into MPS format, i.e.\ $\ket{f} \approx \sum_{\{x_{i,\alpha}\}} A_{1}^{x_{1,1}} A_{2}^{x_{1,2}} \dots A_{n}^{x_{d,n_{\text{x}}}} \ket{x_{1,1}, x_{1,2}, \dots, x_{d,n_{\text{x}}}}$ where $A_{i}^{x_{i,\alpha}}$ is a $\chi_{i} \times \chi_{i+1}$-dimensional matrix and $\chi_{1} = 1 = \chi_{n+1}$.
The so-called bond dimension of the MPS is defined as $\chi = \max_i \chi_i$.
Although a generic state requires $\chi = 2^{\lfloor n / 2 \rfloor}$ for an accurate MPS representation, many practically relevant functions can be represented by MPSs of small $\chi$~\cite{khoromskij2011d,oseledets_2013,lubasch_multigrid,garcia-ripoll_2021,gourianov_2022,ye_2022,ye2024,gourianov_2025}.
The concepts described here for MPSs readily carry over to the more general TTN and CTN ansatzes~\cite{tindall_compressing_2024}.
Throughout our numerical analysis, for the ideal $\ket{f}$ we use a TCI-generated CTN which has negligible approximation errors for the functions considered in the paper (see~\cite{supp} for details).

\textit{IQSP}---As introduced in Fig.~\ref{fig:1}, IQSP prepares the desired multivariate function $\ket{F}$ on a quantum computer with the help of a sequence of $K+1$ smoothly connected functions $\ket{f(\lambda_k)}$ where $k \in \{0, 1, 2, \dots, K\}$ and $\ket{f(\lambda_K)} = \ket{F}$.
We assume that these functions are normalized such that $\mathcal{N} = 1$ and satisfy
\begin{align}\label{eq:smooth_conn}
 \big\Vert{\ket{f(\lambda_k)} - \ket{f(\lambda_{k+1})}}\big\Vert \leq \epsilon
\end{align}
where $k \in \{0, 1, \dots, K-1\}$ and $\epsilon \ll 1$.
The PQC $\ket{\phi(\bm{\theta})} = U(\bm{\theta})\ket{\bm{0}}$ has $M$ variational parameters $\bm{\theta} = (\theta_1, \theta_2, \dots, \theta_M)$.
We assume that the initial PQC for $k = 0$ is easy to prepare.
Then we iterate over $k = 1, 2, \dots, K$ and, for each $\lambda_k$, minimize the infidelity cost function:
\begin{align}\label{eq:infid}
 \mathcal{I}(\bm{\theta}) = 1 - \text{Tr}\Big[{ \rho(\bm{\theta}) \ketbra{f(\lambda_k)}{f(\lambda_k)} }\Big],
\end{align}
where $\rho(\bm{\theta}) = U(\bm{\theta}) \ketbra{\bm{0}}{\bm{0}} U^\dag(\bm{\theta})$ and we initialize the optimization using the final parameters $\bm{\theta}$ of the previous iteration step $k-1$.

Equation~\eqref{eq:smooth_conn} guarantees that, after the optimization at $\lambda_k$, we have a good ansatz to warm-start the optimization at $\lambda_{k+1}$.
To fulfill Eq.~\eqref{eq:smooth_conn}, we need to use a sufficiently small $\delta \lambda = \lambda_{k+1}-\lambda_{k}$:
\begin{align}\label{eq:delta_lambda}
 \delta\lambda \lessapprox \epsilon \norm{\frac{\partial \ket{f(\lambda)}}{\partial \lambda}\Big|_{\lambda_k}}^{-1},
\end{align}
which we prove in~\cite{supp}.
Using this result, Theorem 4 in~\cite{puig_variational_warm_start} can be straightforwardly adapted to establish that the cost function variance has a lower bound that decreases polynomially with $M$ and the barren plateau problem can be avoided.
For functions $\ket{f(\lambda)}$ that are smooth with respect to $\lambda$, we expect that the norm of the first-derivative vector in Eq.~\eqref{eq:delta_lambda} is small such that $\delta \lambda$ does not need to be very small.

In our IQSP implementation, PQCs are optimized by making use of automatic differentiation~\cite{baydin2018automatic} provided by an extended version of the Python torchquantum library~\cite{hanruiwang2022quantumnas}.
For the minimization of the infidelity~\eqref{eq:infid}, we use the Adam optimizer with an initial learning rate $l_{\text{r}} = 10^{-2}$~\cite{kingma2014adam} and $n_{\text{epochs}}$ epochs where each epoch is one Adam optimization step covering all variational parameters once.
The PQCs considered are described in the End Matter and are composed of generic variational SU(4) gates~\cite{wiersema2024herecomessun}.
First we run the IQSP algorithm without taking noise into account.
Then, for the experimental realization on Quantinuum's H2 quantum computer, we perform noise-aware optimization only at the last IQSP step for $\lambda_K$ using the compiled PQC which consists of Quantinuum's native quantum gates~\cite{moses2023race,decross2025}.

We apply IQSP to the preparation of multivariate Gaussians:
\begin{align}\label{eq:normal}
 f(\bm{x}, \lambda; \bm{\mu}, \Sigma) = \frac{1}{\mathcal{N}} \exp\left(-\frac{\lambda}{2}(\bm{x}-\bm{\mu})^{\text{T}} \Sigma^{-1} (\bm{x}-\bm{\mu})\right),
\end{align}
where $\bm{\mu}$ is the vector of means and $\Sigma$ the so-called covariance matrix.
For our analysis, we choose $\mu_i = 0.5 \; \forall i$ and $\Sigma$ to be tridiagonal with entries $s_0$ on the diagonal and $\gamma s_0$ on the off-diagonals.
We fix $s_0 = 0.05$, $\gamma = 0.2$ and $\lambda \in [0, 1]$.
In the following, we use $n_{\text{x}} = 6$ qubits to represent each variable $x_i$, which corresponds to $64$ gridpoints for each dimension.
If a finer resolution is required, one can use larger $n_{\text{x}}$ or Fourier interpolation~\cite{ramos-calderer_efficient_2022} (see~\cite{supp} for additional details).

\begin{figure}
\centering
\begin{overpic}[width=\linewidth]{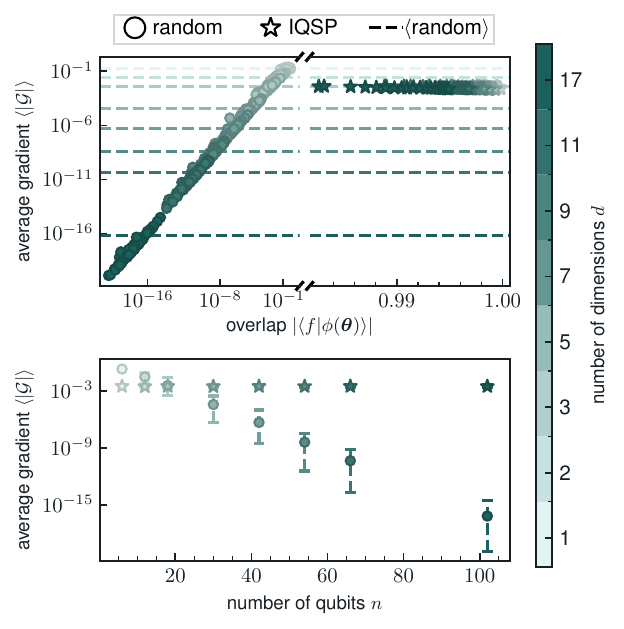}
\put(9,92){(a)}
\put(9,42){(b)}
\end{overpic}
\caption{\label{fig:2}
Gradient magnitudes in IQSP versus random parameter initialization.
(a) Average gradient magnitude~\eqref{eq:gradient} versus overlap at the beginning of the optimization.
Each data point represents a single realization and each dashed line the average over 100 random initializations.
(b) Average gradient magnitude~\eqref{eq:gradient} as a function of the qubit count $n$, whereby the error bars range from the minimum to the maximum gradient magnitude of 100 random parameter choices.
}
\end{figure}

\textit{Gradient magnitudes}---We investigate the absence of barren plateaus in our optimization landscapes by studying the relation between the average gradient magnitude
\begin{align}\label{eq:gradient}
 \langle|\mathcal{G}|\rangle = \frac{1}{M} \sum_{i=1}^M \abs{\frac{\partial \mathcal{I}(\bm{\theta})}{\partial \theta_i}}
\end{align}
and the target state overlap as well as the system size.
We use comb-like PQCs with $L = 3$ layers.
An example of such a PQC with a single layer for $d = 2$ dimensions is shown in Fig.~\ref{fig:1} and with $L = 3$ for $d = 4$ in the End Matter.
For each system size $n$, we randomly initialize the parameters $\bm{\theta}$ and compute both the overlap with the target function and the average gradient magnitude.
This procedure is carried out for several different random parameter choices.
Additionally, we compute the initial overlaps and gradients at each iteration of the IQSP algorithm which we run with $\delta\lambda = 0.05$.
The results are shown in Fig.~\ref{fig:2}.
For randomly initialized PQCs, we observe the expected barren plateau phenomenon that causes gradient magnitudes to decrease exponentially with system size.
In stark contrast, for IQSP, both overlap and gradient magnitudes are large and show no significant system size dependence.
This is numerical evidence that IQSP can avoid the barren plateau problem and efficiently prepare multivariate functions on quantum computers.
The final infidelities of the IQSP-optimized PQCs are reported in~\cite{supp} and, e.g.\ for $d = 17$, we obtain $\mathcal{I} = 4.3 \cdot 10^{-3}$.
In~\cite{supp}, we also perform a similar study on one-dimensional Gaussians for increasing values of $n_{\text{x}}$ up to $n_{\text{x}} = 64$.
In this context, we numerically demonstrate the efficiency of IQSP when higher resolution is desired.

\textit{Noise-aware optimization}---We now explain how experimental noise can be included in IQSP and then optimize PQCs specifically for Quantinuum's H2-2 platform~\cite{moses2023race,decross2025}.
The introduction of noise in IQSP requires working with density matrices $\rho$ instead of pure states.
We employ a density matrix extension of the CTN state as a locally purified TN~\cite{werner2016positive}.
Single-qubit gates are assumed to be noiseless.
We assume that Quantinuum's native two-qubit $Z \otimes Z$ rotation gates of angle $\theta$~\cite{moses2023race,decross2025}, acting on qubits at positions $i$ and $j$, are affected by a quantum noise operator $K_i(\theta) \otimes K_j(\theta)$ where $K_i(\theta)(\rho) = \left[1 - 3r(\theta)\right] \rho + r(\theta) \sum_{\gamma = 1}^{3} \sigma_i^{(\gamma)} \rho \sigma_i^{(\gamma)}$ is a single-qubit depolarizing channel~\cite{nielsen2010quantum}.
Here
\begin{align}
 r(\theta) &= \frac{1}{3} \left(1-\sqrt{1-\frac{5}{4}\epsilon(\theta)}\right)\\
 \epsilon(\theta) &= 2.1 \cdot 10^{-4} + 1.43 \cdot 10^{-3} \theta
\end{align}
are obtained from the device's calibration data and $\sigma_i^{(1)}$, $\sigma_i^{(2)}$ and $\sigma_i^{(3)}$ are the Pauli $X$, $Y$ and $Z$ matrices, respectively, acting on qubit $i$.

\begin{figure}
\includegraphics[width=0.45\textwidth]{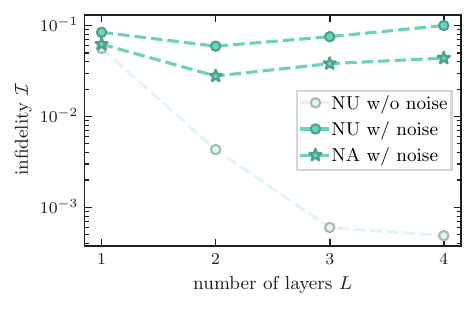}
\caption{\label{fig:3}
Noise-aware IQSP optimization.
Infidelity~\eqref{eq:infid} after noise-unaware (NU) and noise-aware (NA) optimization, evaluated with (w/) and without (w/o) noise, for a four-dimensional Gaussian realized using 24 qubits.
}
\end{figure}

We apply noise-aware IQSP to the preparation of a Gaussian with $d = 4$ and $n = 24$.
We analyze the effect of noise on the infidelity~\eqref{eq:infid} for different numbers of layers $L \in\{1, 2, 3,4\}$ after $n_{\text{epochs}} = 10^4$ optimization epochs.
We remove two-qubit gates with angles $\theta \leq 10^{-4}$ since the constant error introduced by their presence is higher than their rotation angle.
In Fig.~\ref{fig:3} we show the infidelity as a function of the number of layers.
We see that the noise-aware approach significantly improves the final results over the ones of the noise-unaware optimization.
In the presence of noise, the best infidelity reads $\mathcal{I} = 0.028$ and is obtained for $L = 2$.

\begin{figure}
\centering
\includegraphics[width=\columnwidth]{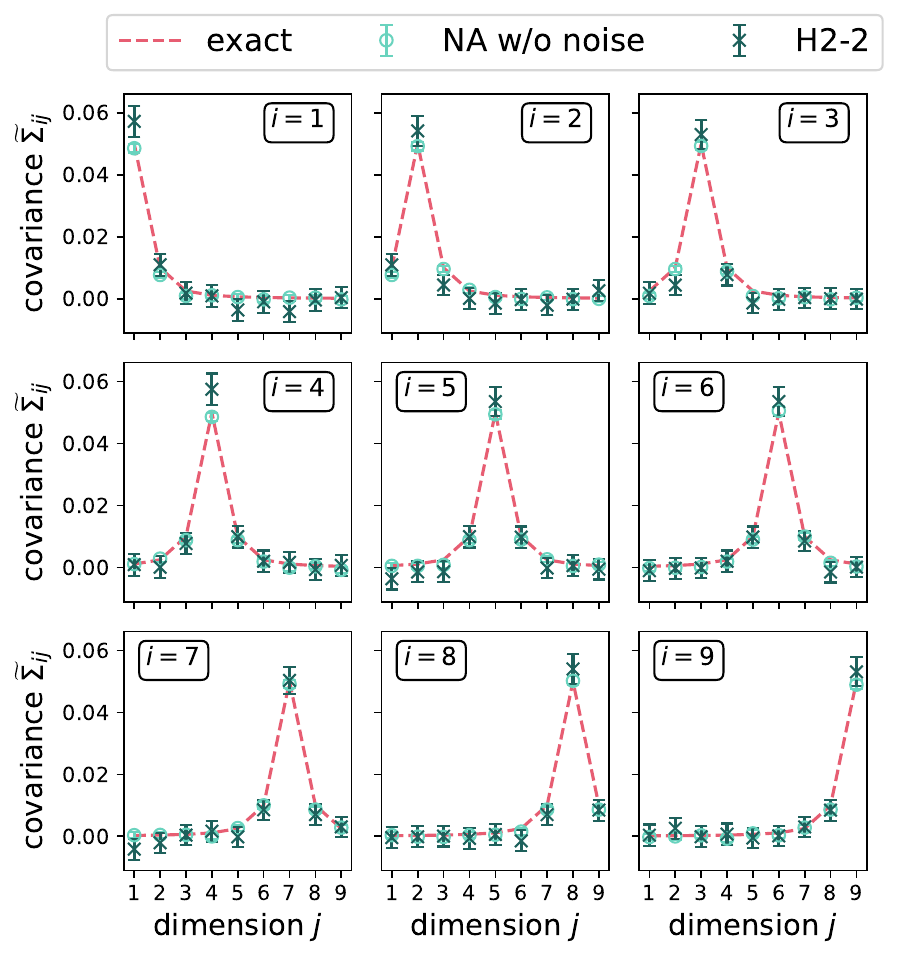}
\caption{\label{fig:4}
Covariances~\eqref{eq:covs} measured on H2-2.
We create the 9-dimensional Gaussian with covariance matrix~\eqref{eq:gaus_quadr} using $54$ qubits.
The PQC is obtained by noise-aware (NA) optimization and is simulated without (w/o) noise and run on H2-2.
The error bars represent 2 standard deviations.
The noiseless results (empty circles) are estimated using $n_{\text{shots}} = 10^4$ and the experimental results on H2 (crosses) using $n_{\text{shots}} = 1024$.
}
\end{figure}

\textit{Experimental demonstration}---In this section, we report on the preparation of a Gaussian with $d = 9$ using $n = 54$ qubits on Quantinuum's H2-2 quantum computer.
Motivated by our previous results for $d = 4$, here we fix $L = 2$.
We consider the Gaussian with the covariance matrix:
\begin{align}\label{eq:gaus_quadr}
 \Sigma_{i j} = \frac{\gamma s_0}{\abs{i-j}^2}\; \forall\; i \neq j, \quad \Sigma_{i i} = s_0,
\end{align}
where $i, j \in \{1, 2, \dots, 9\}$.
Using noise-unaware IQSP, we reach an infidelity $\mathcal{I} = 0.01$ evaluated without noise.
We perform noise-aware optimization and improve $\mathcal{I} = 0.13$ evaluated in the presence of noise to $\mathcal{I} = 0.044$.
The procedure reduces the two-qubit gate count by $\approx 20\%$, from $318$ to $255$ gates.
We run the optimized PQC on Quantinuum hardware $n_{\text{shots}}$ times and compute the covariances:
\begin{equation}\label{eq:covs}
 \widetilde{\Sigma}_{i j} = \mathbb{E}[x_i x_j] - \mathbb{E}[x_i] \mathbb{E}[x_j],
\end{equation}
where $\mathbb{E}[\mathcal{O}]$ is the average value of $\mathcal{O}$ across $n_{\text{shots}}$ measurements of $\mathcal{O}$.
Error bars are calculated with the help of the variance:
\begin{equation}
 \text{Var}[\widetilde{\Sigma}_{i j}] = \frac{1}{n_{\text{shots}}-1} \left(\widetilde{\Sigma}_{i j}^2 + \widetilde{\Sigma}_{i i} \widetilde{\Sigma}_{j j}\right),
\end{equation}
via its square root, i.e.\ the standard deviation.
In Fig.~\ref{fig:4}, we compare the experimental results to the exact covariance matrix and to results obtained from the noiseless simulation of the PQC using the qmatchtatea MPS emulator~\cite{qmatchatea}.
The emulator is used with a sufficiently large MPS bond dimension such that its results are numerically exact.
We conclude from Fig.~\ref{fig:4} that IQSP produces an accurate representation of the 9-dimensional Gaussian on hardware.
Additional experimental results, including all measurement data for mean values and covariances, are provided in~\cite{supp}.

\textit{Conclusions and outlook}---In this paper, we introduce IQSP and show that it can solve the crucial problem of how to efficiently realize high-dimensional multivariate functions on current quantum computers.
Because QSP is needed in many quantum algorithms, IQSP paves the way for various new applications of quantum computing hardware that is already available today.

Next, it is important to assess the performance of IQSP on the preparation of other functions, such as with long-range interactions between the variables, and compare it to alternative QSP approaches.
We make a first step in this direction in the End Matter where we compare IQSP with the QSP algorithm of~\cite{rosenkranz_2025}.

Another interesting research direction is the formulation of a new variant of IQSP that does not require an efficient TN description of the function that is supposed to be prepared on a quantum computer.
In the End Matter, we propose circuit cross interpolation (CCI) as a new tool that reformulates TCI for the optimization of generic quantum circuits, by evaluating them on an incrementally built set of computational basis states, so-called pivots.
The set of pivots together with the corresponding function values represent the function, and they enable the preparation of the quantum state without requiring the fidelity.
By avoiding the fidelity calculation, CCI does not rely on TN at all and neither does IQSP based on CCI.
While this approach will likely lead to a broader applicability of IQSP, we find that, for the problems considered in this paper, the TN-based IQSP version presented in the main text is more efficient.

\textit{Acknowledgments}---We thank Frederic Sauvage, Pranav Kalidindi, and Conor Mc Keever for useful discussions.
We are also grateful to Lewis Wright and Guillermo Preisser for carefully reading the manuscript and providing helpful feedback.

\bibliography{references}

\onecolumngrid
\vspace{1cm}
\begin{center}
\textbf{\large End Matter}    
\end{center}
\twocolumngrid

\begin{figure}[t]
\centering
\begin{overpic}[width=0.9\linewidth]{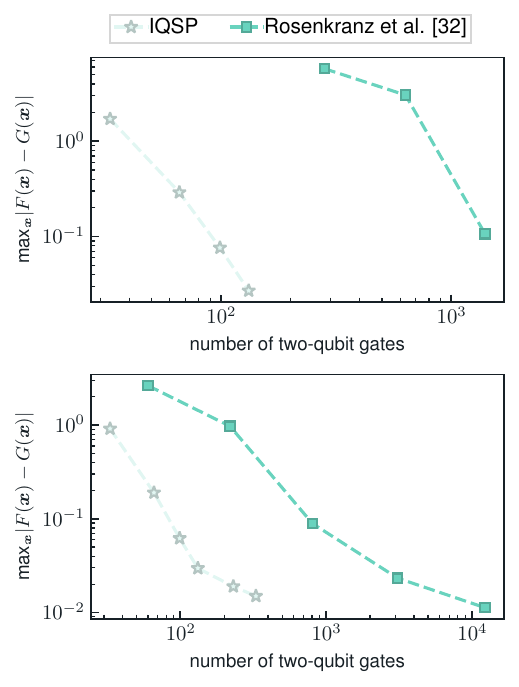}
\put(9,93){(a)}
\put(9,47){(b)}
\end{overpic}
\caption{\label{fig:5}
Error~\eqref{eq:err_max} as a function of two-qubit gate count.
(a) We consider the Ricker wavelet and compare to the Chebyshev polynomial approach of~\cite{rosenkranz_2025}.
(b) We consider the Student's t-distribution and compare to the Fourier series approach of~\cite{rosenkranz_2025}.
}
\end{figure}

\textit{Comparison to prior work}---Here we want to compare the performance of IQSP with the one of the QSP approach~\cite{rosenkranz_2025}.
To that end, we apply IQSP to the preparation of the Ricker wavelet and Student's t-distribution considered in~\cite{rosenkranz_2025}.
The Ricker wavelet is defined as:
\begin{align}\label{eq:ricker}
\begin{split}
 F(\bm{x}; \bm{\mu}, \sigma) =& \frac{1}{\mathcal{N}} \left[1-\frac{(\bm{x}-\bm{\mu})^\text{T}(\bm{x}-\bm{\mu})}{2\sigma^2} \right]\\
 & \times \exp[-\frac{(\bm{x}-\bm{\mu})^\text{T}(\bm{x}-\bm{\mu})}{2\sigma^2}],
\end{split}
\end{align}
and the Student's t-distribution as:
\begin{align}\label{eq:student}
 F(\bm{x}; \bm{\mu}, \Sigma) = \frac{1}{\mathcal{N}} \left[1 + (\bm{x}-\bm{\mu})^\text{T}\Sigma^{-1}(\bm{x}-\bm{\mu}) \right]^{-\frac{3}{2}}.
\end{align}
To be consistent with~\cite{rosenkranz_2025}, we set $d = 2$, $n_{\text{x}} = 6$, $\sigma = 0.25$, $\mu_i = 0.5\; \forall i$, $\Sigma_{i i} = 0.05\; \forall i$, and $\Sigma_{i j} = 0\; \forall\; i \neq j$.
We run IQSP using $\lambda \in [0, 1]$, $\delta\lambda = 0.05$, $n_{\text{epochs}} = 10^3$ for $\lambda < 1$ and $n_{\text{epochs}} = 10^4$ for $\lambda = 1$.
We calculate the error:
\begin{align}\label{eq:err_max}
 \epsilon_{\max} = \max_{\bm{x}} |F(\bm{x}) - G(\bm{x})|,
\end{align}
which is also used in~\cite{rosenkranz_2025}, where $F(\bm{x})$ is the desired target function and $G(\bm{x}) = \sqrt{\sum_{\bm{x}} \abs{F(\bm{x})}^2} \braket{\bm{x}}{\phi(\bm{\widetilde{\theta}})}$ is the rescaled quantum state $\ket{\phi(\bm{\theta})}$ created by the optimized PQC with the optimal parameters $\bm{\theta} = \bm{\widetilde{\theta}}$.
Figure~\ref{fig:5} shows our results.
We see that, compared with~\cite{rosenkranz_2025}, IQSP reaches the same errors using approximately one order of magnitude fewer two-qubit gates.

\textit{IQSP using circuit cross interpolation}---In this section, we explore an alternative IQSP approach that does not use TNs.
We recall that each iteration of the IQSP algorithm is an optimization problem and requires the efficient evaluation of a cost function.
In the main text, we evaluate the cost function via a TN contraction which makes use of the target function representation by a TCI-generated CTN.
Our goal is to extend the concepts underlying TCI to directly interpolate the target function using a PQC.
We refer to the procedure as circuit cross interpolation (CCI).

Analogously to TCI, we want to choose, in an optimal way, a relatively small number of function values corresponding to certain function input values, so-called pivots, to capture the function behavior and optimize the variational PQC parameters.
We achieve this by growing and modifying the set of pivots $P = \{\bm{p_i}\}_{i = 1, 2, \dots, n_P}$ during the optimization using function input values for which the PQC approximation errors are maximum such that the subsequent optimization steps try to correct the largest errors previously observed.

Specifically, the CCI method works as follows:
\begin{enumerate}
\item Initialize $P$ with $\bm{p_1} = \text{argmax}_{\bm{x}} f(\bm{x})$ if the maximum is known, otherwise select the first pivot randomly.
\item Optimize the PQC $\ket{\phi(\bm{\theta})}$ to minimize the cost function:
\begin{align}
 \mathcal{C} = \sum_{i=1}^{n_P} \left|f(\bm{p_i}) - \braket{\bm{p_i}}{\phi(\bm{\theta})}\right|^2.
\end{align}
\item Create a new set of pivots $\widetilde{P}$, e.g., by performing local bit modifications on pivots in $P$.
For example, if $n = 3$ and we have only one pivot, we can create the following new pivots:
\begin{align}
 \{100\} \rightarrow \{\textcolor{teal}{01}0, \textcolor{teal}{0}00, 1\textcolor{teal}{1}0, 10\textcolor{teal}{1}, 1\textcolor{teal}{11}\},
\end{align}
where we highlight in green the bits that are modified.
From all the new pivots, choose the one pivot that, when added to $P$, gives the largest cost function value $\mathcal{C}$ that can be obtained by adding exactly one pivot of $\widetilde{P}$ to $P$.
Enlarge $P$ by this pivot.
\item Repeat steps 2 and 3 above until the desired maximum number of pivots is reached.
Then continue repeating 2 and 3 with fixed number of pivots which are modified in each iteration.
Repeat until the desired convergence criterion is met.
\end{enumerate}

We have benchmarked this approach on the preparation of one-dimensional Gaussians and confirmed that it can be an accurate QSP algorithm.
For the multivariate functions studied in the paper, however, we have found that using TCI-generated CTNs in IQSP is more efficient than CCI-based QSP.

\begin{figure}
\centering
\includegraphics[width=\linewidth]{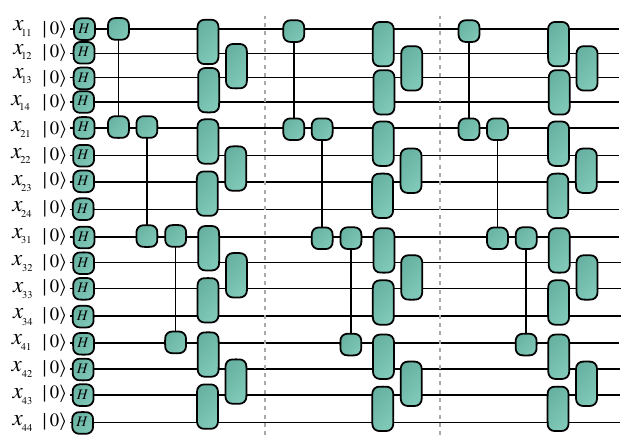}
\caption{\label{fig:6}
Comb-like PQC used in the paper.
We show the PQC ansatz of $L = 3$ layers for $d = 4$ dimensions with $n_{\text{x}} = 4$ qubits per dimension.
The boxes labelled $H$ represent Hadamard gates and the other boxes are variational SU(4) gates.
The dashed gray lines separate the 3 different layers.
}
\end{figure}

\textit{Explicit circuit ansatz}---Figure~\ref{fig:6} depicts a specific example of the comb-like PQC architecture that is used throughout the paper, before it is compiled into hardware-native gates.
Every qubit is initialized in the state $\ket{0}$.
Then a single layer of Hadamard gates is applied, which realizes a normalized constant function on the quantum computer.
This is followed by $L = 3$ layers of gates, where in each layer we have one set of gates that connects qubits corresponding to different dimensions in a staircase-like construction and another set of gates, connected in a brickwork-like fashion, that entangles the qubits associated with each individual dimension.

\end{document}


\title{Supplemental Material:\\
Efficient quantum state preparation of multivariate functions using tensor networks}

\author{Marco Ballarin\orcMarco}\email{marco.ballarin@quantinuum.com}\affA
\author{Juan Jos\'{e} Garc\'{i}a-Ripoll\orcJuanjo}\affB
\author{David Hayes\orcDave}\affC
\author{Michael Lubasch\orcLubi}\email{michael.lubasch@quantinuum.com}\affA

\date{November 19, 2025}

\maketitle

We analyze the accuracy of TCI in Sec.~\ref{sec:tci}.
In Sec.~\ref{sec:low_bound}, we derive a lower bound on the IQSP cost function variance.
Section~\ref{sec:resolution} contains a study of the performance of IQSP when higher resolution is desired.
We investigate the accuracy of IQSP in Sec.~\ref{sec:infids}.
Section~\ref{sec:exp_detail} provides experimental details.

\section{Accuracy of TCI}
\label{sec:tci}

\begin{figure}
\centering
\includegraphics[width=0.8\linewidth]{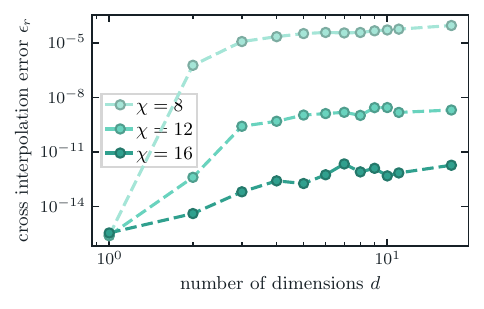}
\caption{\label{fig:tci_err}
Average relative error~\eqref{eq:tci_err} of the TCI-generated CTN approximations of the multivariate Gaussians studied in the main text as a function of the dimensionality.
In the calculation of~\eqref{eq:tci_err}, we randomly choose $n_\text{avg} = 10^4$ function input values $\bm{x_i}$.
}    
\end{figure}

Here, we numerically demonstrate that TCI-generated CTNs are accurate representations of the multivariate Gaussians considered in the paper.
We use CTNs of bond dimensions up to $\chi = 16$ and prepare the multivariate Gaussians defined in Eq.~(6) of the main text for several values of $d$.
To quantify the error of the CTN approximation, we compute the average relative difference between the CTN $\ket{\psi_{\text{CTN}}}$ and the target function $f$ evaluated for $n_{\text{avg}}$ function input values $\bm{x_i}$:
\begin{align}\label{eq:tci_err}
 \epsilon_r = \frac{1}{n_\text{avg}} \sum_{i=1}^{n_\text{avg}} \abs{\frac{f(\bm{x_i}) - \braket{\bm{x_i}}{\psi_{\text{CTN}}}}{f(\bm{x_i})}}.
\end{align}
Figure~\ref{fig:tci_err} shows the results.
We see that, for $\chi = 16$, the largest error is $\approx 10^{-12}$.

\section{Lower bound on the IQSP cost function variance}
\label{sec:low_bound}

Our goal is to derive Eq.~(5) of the main text which, combined with~\cite{puig_variational_warm_start}, leads to a lower bound on the IQSP cost function variance.

We recall that Eq.~(5) follows from Eq.~(3) of the main text:
\begin{align}
 \big\Vert{\ket{f(\lambda_k)} - \ket{f(\lambda_{k+1})}}\big\Vert \leq \epsilon.
\end{align}
We assume that $\ket{f(\lambda)}$ is real-valued and normalized such that
\begin{align}\label{eq:eq3}
 \big\Vert{\ket{f(\lambda_k)} - \ket{f(\lambda_{k+1})}}\big\Vert = \sqrt{2 - 2 \braket{f(\lambda_k)}{f(\lambda_{k+1})}}.
\end{align}
Let us use a Taylor expansion:
\begin{align}\label{eq:taylor}
\begin{split}
 \ket{f(\lambda_{k+1})} &= \ket{f(\lambda_k+\delta\lambda)}\\
 &= \ket{f(\lambda_k)} + \ket{f'(\lambda_k)} \delta\lambda\\
 & \quad + \ket{f''(\lambda_k)} \frac{(\delta\lambda)^{2}}{2} + O[(\delta\lambda)^3],
\end{split}
\end{align}
where $\ket{f'(\lambda_k)}$ ($\ket{f''(\lambda_k)}$) is the vector of first (second) derivatives of $\ket{f(\lambda)}$ with respect to $\lambda$ evaluated at $\lambda_k$.
The following identities are true:
\begin{align}
 \braket{f(\lambda)}{f(\lambda)} &= 1,\\
 \braket{f(\lambda)}{f'(\lambda)} &= \frac{1}{2} \frac{\partial}{\partial \lambda} \braket{f(\lambda)}{f(\lambda)} = 0,\\
 \braket{f(\lambda)}{f''(\lambda)} &= -\norm{\ket{f'(\lambda)}}^2.
\end{align}
These identities together with the Taylor expansion~\eqref{eq:taylor} enable us to approximate Eq.~\eqref{eq:eq3} for sufficiently small $\delta \lambda$ by:
\begin{align}
 \sqrt{2 - 2 \braket{f(\lambda_k)}{f(\lambda_{k+1})}} \approx \norm{\ket{f'(\lambda_k)}} \delta \lambda,
\end{align}
from which Eq.~(5) of the main text readily follows.

\section{Higher resolution}
\label{sec:resolution}

\begin{figure}
\centering
\includegraphics[width=\linewidth]{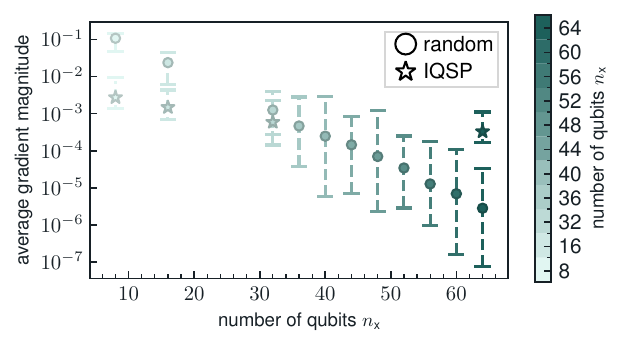}
\caption{\label{fig:1d_gauss}
Gradient magnitudes in IQSP versus random parameter initialization for one-dimensional Gaussians.
We show the average initial gradient magnitude as a function of qubit count.
Each data point corresponds to the average over 1000 random initializations.
The error bars range from the minimum and the maximum of the values.
}
\end{figure}

In the main text, we use $n_{\text{x}} = 6$ qubits for each dimension of the multivariate Gaussians.
While a grid of $2^{n_{\text{x}}} = 64$ grid points per dimension can often be sufficient to faithfully represent multivariate Gaussians, a higher resolution might be desirable in certain cases.
There exist previous studies in which one-dimensional Gaussians are prepared using up to $100$ qubits~\cite{melnikov_quantum_2023,bohun_scalable_2024}.
The IQSP algorithm can also prepare such functions with extraordinarily high resolution.
Figure~\ref{fig:1d_gauss} shows that average gradient magnitudes remain large during IQSP as we increase $n_{\text{x}}$, whereas random parameter initialization leads to a significantly faster decrease of average gradient magnitudes with $n_{\text{x}}$.

We note that, to increase the resolution, an interesting alternative approach is Fourier interpolation~\cite{ramos-calderer_efficient_2022}.

\section{Accuracy of IQSP}
\label{sec:infids}

\begin{figure}
\centering
\includegraphics[width=0.8\linewidth]{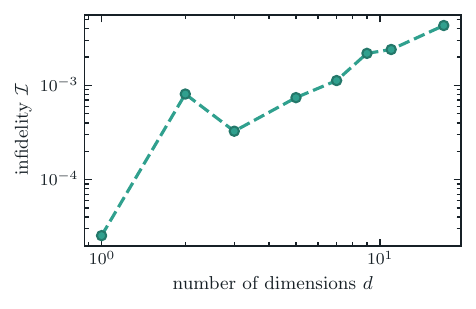}
\caption{\label{fig:infid_scaling}
Infidelity after IQSP optimization for the $d$-dimensional Gaussians of the main text.
We optimize comb-like PQCs and run the IQSP algorithm with $\delta\lambda = 0.05$ for $\lambda \in [0, 1]$.
}
\end{figure}

In the main text, our main focus is on fidelities and gradient magnitudes at the beginning of the IQSP optimization iterations.
Figure~\ref{fig:infid_scaling} shows how the final infidelity after the IQSP optimization depends on the dimensionality $d$ of the Gaussians considered in the main text.
We observe that, even for $d = 17$, using $n = 102$ qubits, IQSP leads to the remarkably low infidelity of $\mathcal{I} = 4.3 \cdot 10^{-3}$.

\vspace*{5mm}

\section{Experimental details}
\label{sec:exp_detail}

Here, we collect experimental details on the preparation of the 9-dimensional Gaussian studied in the main text.
The experiment was run on the 29th of June 2025.
Table~\ref{tab:means} contains the measured mean values of the variables and Tab.~\ref{tab:cov} the measured covariances.

\begin{table}[b]
\centering
\begin{tabular}{c|ccc}
 dimension $d$\;\; & H2-2 & noiseless & exact \\\hline
 $\langle x_1\rangle$ & \;$0.499\pm0.007$\;&\; $0.502\pm0.002$ \;& $0.5$\\
 $\langle x_2\rangle$ & $0.499\pm0.007$ & $0.500\pm0.002$ & $0.5$ \\
 $\langle x_3\rangle$ & $0.492\pm0.007$ & $0.501\pm0.002$ & $0.5$\\
 $\langle x_4\rangle$ & $0.503\pm0.007$ & $0.501\pm0.002$ & $0.5$\\
 $\langle x_5\rangle$ & $0.505\pm0.007$ & $0.501\pm0.002$ & $0.5$\\
 $\langle x_6\rangle$ & $0.490\pm0.007$ & $0.499\pm0.002$ & $0.5$\\
 $\langle x_7\rangle$ & $0.508\pm0.007$ & $0.501\pm0.002$ & $0.5$\\
 $\langle x_8\rangle$ & $0.486\pm0.007$ & $0.499\pm0.002$ & $0.5$\\
 $\langle x_9\rangle$ & $0.489\pm0.007$ & $0.499\pm0.002$ & $0.5$\\
\end{tabular}
\caption{\label{tab:means}
Mean values of all variables for the $9$-dimensional Gaussian considered.
The values are estimated via $n_{\text{shots}} = 1024$ on H2-2 and $n_{\text{shots}} = 10^4$ in the noiseless simulation.
}
\end{table}

\begin{table*}
\centering
\resizebox{\textwidth}{!}{
\begin{tabular}{c|ccccccccc}
 dimension $d$\;\; & $x_1$ & $x_2$ & $x_3$ & $x_4$ & $x_5$ & $x_6$ & $x_7$ & $x_8$ & $x_9$\\\hline
 $ x_1$ & $0.057\pm0.003$ & $0.011\pm0.002$ & $0.002\pm0.002$ & $0.001\pm0.002$ & $-0.004\pm0.002$ & $-0.001\pm0.002$ & $-0.004\pm0.002$ & $-0.0\pm0.002$ & $0.0\pm0.002$\\

 $ x_2$ & $0.011\pm0.002$ & $0.054\pm0.002$ & $0.004\pm0.002$ & $0.0\pm0.002$ & $-0.001\pm0.002$ & $-0.0\pm0.002$ & $-0.002\pm0.002$ & $-0.0\pm0.002$ & $0.003\pm0.002$\\

 $x_3$ & $0.002\pm0.002$ & $0.004\pm0.002$ & $0.053\pm0.002$ & $0.008\pm0.002$ & $-0.001\pm0.002$ & $-0.0\pm0.002$ & $0.0\pm0.002$ & $0.0\pm0.002$ & $-0.0\pm0.002$\\

 $x_4$ & $0.001\pm0.002$ & $0.0\pm0.002$ & $0.008\pm0.002$ & $0.058\pm0.003$ & $0.01\pm0.002$ & $0.002\pm0.002$ & $0.002\pm0.002$ & $-0.001\pm0.002$ & $0.001\pm0.002$\\

 $x_5$ & $-0.004\pm0.002$ & $-0.001\pm0.002$ & $-0.001\pm0.002$ & $0.01\pm0.002$ & $0.054\pm0.002$ & $0.01\pm0.002$ & $-0.0\pm0.002$ & $0.001\pm0.002$ & $-0.001\pm0.002$\\

 $x_6$ & $-0.001\pm0.002$ & $-0.0\pm0.002$ & $-0.0\pm0.002$ & $0.002\pm0.002$ & $0.01\pm0.002$ & $0.054\pm0.002$ & $0.009\pm0.002$ & $-0.001\pm0.002$ & $0.0\pm0.002$\\

 $x_7$ & $-0.004\pm0.002$ & $-0.002\pm0.002$ & $0.0\pm0.002$ & $0.002\pm0.002$ & $-0.0\pm0.002$ & $0.009\pm0.002$ & $0.05\pm0.002$ & $0.007\pm0.002$ & $0.003\pm0.002$\\

 $x_8$ & $-0.0\pm0.002$ & $-0.0\pm0.002$ & $0.0\pm0.002$ & $-0.001\pm0.002$ & $0.001\pm0.002$ & $-0.001\pm0.002$ & $0.007\pm0.002$ & $0.054\pm0.002$ & $0.008\pm0.002$\\

 $x_9$ & $0.0\pm0.002$ & $0.003\pm0.002$ & $-0.0\pm0.002$ & $0.001\pm0.002$ & $-0.001\pm0.002$ & $0.0\pm0.002$ & $0.003\pm0.002$ & $0.008\pm0.002$ & $0.053\pm0.002$\\
\end{tabular}}
\caption{\label{tab:cov}
Covariance matrix elements measured on H2-2.
We consider the $9$-dimensional Gaussian defined in the main text.
Each data point is calculated using $n_{\text{shots}} = 1024$ measurements.
}
\end{table*}

\bibliography{references}